\documentclass[fleqn,twoside]{article}
\usepackage{espcrc2}

\usepackage{graphicx}
\usepackage{bm}

\def\la{\mathrel{\mathpalette\fun <}}
\def\ga{\mathrel{\mathpalette\fun >}}
\def\fun#1#2{\lower3.6pt\vbox{\baselineskip0pt\lineskip.9pt
  \ialign{$\mathsurround=0pt#1\hfil##\hfil$\crcr#2\crcr\sim\crcr}}}

\newcommand{\D}{{\rm d}}

\begin{document}

\title{Collective flavor transitions of supernova neutrinos}

\author{G\"unter Sigl\address[HAM]{II. Institut f\"ur theoretische Physik, Universit\"at Hamburg,
 Luruper Chaussee 149, D-22761 Hamburg, Germany},
  Ricard Tom\`as\addressmark[HAM],
 Andreu Esteban-Pretel\address[AHEP]{AHEP Group, Institut de F\'\i sica
 Corpuscular, CSIC/Universitat de Val\`encia,\\
 Edifici Instituts d'Investigaci\'o, Apt.\ 22085,
 46071 Val\`encia, Spain}, Sergio Pastor\addressmark[AHEP],
 Alessandro Mirizzi\address[MPI]{Max-Planck-Institut f\"ur Physik
 (Werner-Heisenberg-Institut), F\"ohringer Ring 6,\\ 80805 M\"unchen,
 Germany}, Georg~G.~Raffelt\addressmark[MPI],
 Pasquale D.~Serpico\address[CERN]{Physics Department, Theory Division, CERN, CH-1211 Geneva 23, Switzerland}}

\begin{abstract}
We give a very brief overview of collective effects in neutrino oscillations in core
collapse supernovae where refractive effects of neutrinos on themselves can
considerably modify flavor oscillations, with possible repercussions for future
supernova neutrino detection. We discuss synchronized and bipolar oscillations, the role
of energy and angular neutrino modes, as well as three-flavor effects. We close with a short
summary and some open questions.
\vspace{1pc}
\end{abstract}

\maketitle


\section{Introduction}
Neutrinos interact via the weak interactions and can thus cause a refractive
effect on each other. Under most circumstances, the resulting self-interaction
potential is much smaller than the vacuum oscillation term or the potential
induced by ordinary matter. However, in the early cooling phase of core collapse
supernovae, the density of neutrinos streaming off the hot nascent neutron star
is sufficiently high to cause non-linear phenomena that can have practical
importance. This was realized only recently in a series of papers of which we
here cite only a
few~\cite{Sawyer:2005jk,Duan:2005cp,Duan:2006an,Hannestad:2006nj,Raffelt:2007yz,EstebanPretel:2007ec,Raffelt:2007cb,Raffelt:2007xt,Duan:2007bt,Fogli:2007bk,EstebanPretel:2007yq,EstebanPretel:2008ni,Fogli:2008pt,Fogli:2008fj}.

To describe collective effects it is useful to describe the neutrinos with
flavor density matrices for each momentum mode. They are defined in terms of the
annihilation operators $a_i$ for neutrinos and $\bar a_j$ for antineutrinos in a
given momentum mode ${\bf p}$~\cite{Dolgov:1980cq,Sigl:1992fn,mckellar&thomson},
\begin{equation}\label{eq:densitymatrixdefinition}
 (\varrho_{\bf p})_{ij}=
 \langle a^\dagger_{i} a_{j}\rangle_{\bf p}
 \hbox{\quad and\quad}
 (\bar \varrho_{\bf p})_{ij}=
 \langle \bar a^\dagger_{j}\,\bar a_{i}\rangle_{\bf p}\,.
\end{equation}
their equations of motion can be written as a commutator of an effective
Hamiltonian with the density matrices,
\begin{eqnarray}\label{eq:eom}
 \partial_t\varrho_{\bf p}&=&-i\biggl[\Omega_{\bf p}+\sqrt{2}\,G_{\rm F}L+\\
  &&\hskip-1cm+\sqrt{2}\,G_{\rm F}\int\!\frac{\D^3{\bf q}}{(2\pi)^3}
 \left(\varrho_{\bf q}-\bar\varrho_{\bf q}\right)
 (1-\cos\theta_{\bf pq}),\varrho_{\bf p}\biggr]\,.\nonumber
\end{eqnarray}
Here, $G_{\rm F}$ is the Fermi constant, in the mass basis the
matrix of vacuum oscillation frequencies for relativistic neutrinos is
$\Omega_{\bf p}={\rm diag}(m_1^2,m_2^2,m_3^2)/2p$ with $p=|{\bf p}|$,
and in the flavor basis $L={\rm diag}(n_e,n_\mu,n_\tau)$,
where $n_e,n_\mu,n_\tau$ are the charged lepton number densities (particle
minus antiparticle density) responsible for the matter induced
potential. When neutrino self-interactions are negligible, a compensation
of the vacuum and matter induced terms in the diagonal part of the Hamiltonian
in Eq.~(\ref{eq:eom}) leads to the well known Mikheyev-Smirnov-Wolfenstein (MSW)
effect~\cite{Wolfenstein:1977ue,Mikheyev:1985aa}.
An equation analogous to Eq.~(\ref{eq:eom}) holds for antineutrinos
with the sign of the vacuum term $\Omega_{\bf p}$ reversed. The last
term in Eq.~(\ref{eq:eom}) describes forward scattering on the neutrino
background where the isotropic density term can lead to self-maintained
coherence, whereas the flux term proportional to the cosine of the
angle $\theta_{\bf pq}$ between momenta ${\bf p}$ and ${\bf q}$ can
lead to self-induced decoherence. The flux of charged leptons has been assumed
to vanish.

In the two-flavor case it is convenient to parametrize vacuum, matter and
self-interaction terms by the frequencies
\begin{eqnarray}\label{eq:para}
  \omega_p&=&\frac{|\Delta m^2|}{2p}\,;\nonumber\\
  \lambda(r)&=&\sqrt2\,G_{\rm F}n_e(r)\,;\\
  \mu(r)&=&\sqrt2\,G_{\rm F}\left[F_{\bar\nu_e}(r)-F_{\bar\nu_x}(r)\right]
  \left\langle1-\cos\theta_{\bf pq}\right\rangle\,,\nonumber
\end{eqnarray}
with $F_i$ are the fluxes of the relevant neutrino species $i$, and $\nu_x$
stands for any one of $\nu_\mu$, $\nu_\tau$, $\bar\nu_\mu$ or $\bar\nu_\tau$. The oscillation
phenomena discussed in the following depend crucially on the relative
size of these three frequencies.

\section{Synchronized and Bipolar Oscillations}\label{sec2}
Since muon and tau neutrinos behave equally in core collapse supernovae up
to an effective $\mu\tau$ potential briefly discussed in Sect.~\ref{sec4},
neutrino oscillations can be described in two-flavor approximation. The
oscillations are then driven by the atmospheric mass squared difference,
$|\Delta m^2_{\rm atm}|=|m_3^2-m_2^2|=2.40^{+0.12}_{-0.11}\times10^{-3}\,{\rm eV}^2$
and the angle $\sin\theta_{13}\le0.04$~\cite{Schwetz:2008er}. For these
parameters, $\omega_p\simeq0.4\,{\rm km}^{-1}$ for a typical neutrino energy
of 15 MeV. Both normal and
inverted hierarchy are still allowed for this case.

Considering only a single two-flavor mode ${\bf p}$ one can expand into
Pauli matrices $\bm\sigma$,
$\varrho=(f+{\bf P}\cdot\bm{\sigma})/2$, analogously for $\bar\varrho$,
$L=(n_0+n_e{\bf L}\cdot\bm{\sigma})/2$, and
$\omega=\left[\omega_0+|\Delta m^2|{\bf B}\cdot\bm{\sigma}/(2p)\right]/2$. With
the vacuum mixing angle $\theta$, ${\bf B}=(\sin2\theta,0,-\cos2\theta)$, the
radial flavor evolution (we use units in which the speed of light is unity)
can be described by~\cite{Hannestad:2006nj}
\begin{equation}\label{eq:onemode}
  \partial_r{\bf P}=\left[+\omega{\bf B}+\lambda(r){\bf L}+\mu(r)
  ({\bf P}-{\bf\bar P})\right]\times{\bf P}\,,
\end{equation}
and an analogous equation for ${\bf\bar P}$ with the sign of $\omega$ reversed.
It is useful to define the vectors
\begin{eqnarray}\label{eq:QD}
  {\bf Q}&\equiv&{\bf P}+{\bf\bar P}-\frac{\omega}{\mu}{\bf B}\,;
  \quad{\bf q}\equiv\frac{\bf Q}{Q}\nonumber\\
  {\bf D}&\equiv&{\bf P}-{\bf\bar P}\,.
\end{eqnarray}
In the absence of matter, $\lambda=0$, and for slowly varying $\mu$, the equations of motion
(now written in terms of time derivatives) are
\begin{equation}\label{eq:onemode2}
  \dot{\bf Q}=\mu{\bf D}\times{\bf Q}\,;\quad\dot{\bf D}=\omega{\bf B}\times{\bf Q}\,,
\end{equation}
which implies
\begin{eqnarray}\label{eq:onemode3}
  |{\bf Q}|={\rm const}\,;&&\quad\sigma\equiv{\bf D}\times{\bf Q}={\rm const}\nonumber\\
  {\bf D}&=&\frac{{\bf q}\times\dot{\bf q}}{\mu}+\sigma{\bf q}\,.
\end{eqnarray}
These are the equations for a spinning top of spin $\sigma$, angular momentum
$({\bf q}\times\dot{\bf q})/\mu$ and moment of inertia $I=ml^2=\mu^{-1}$. Its
energy then has a kinetic and a potential part,
\begin{eqnarray}\label{eq:energy}
  E&=&E_{\rm kin}+E_{\rm pot}=\frac{\mu}{2}{\bf D}^2+\omega({\bf B}\cdot{\bf Q}+Q)\nonumber\\
  &&=\frac{\dot{\bf q}^2}{2\mu}+\frac{\mu}{2}\sigma^2+\omega({\bf B}\cdot{\bf Q}+Q)\,.
\end{eqnarray}
This description is a good approximation in the regime where neutrinos are freely streaming.
The supernova neutrino (number) fluxes are thought to obey the hierarchy $F_{\nu_e}>F_{\bar\nu_e}>F_{\nu_x}$.
One can then adopt initial conditions at the neutrino sphere are typically ${\bf P}=(0,0,1+\varepsilon)$,
$\bar{\bf P}=(0,0,1)$, where the asymmetry parameter in terms of the fluxes of different
neutrino flavors is
\begin{equation}\label{eq:epsdefine}
 \epsilon=
 \frac{F_{\nu_e}-F_{\nu_x}}{F_{\bar\nu_e}-F_{\bar\nu_x}}-1
 =\frac{F_{\nu_e}-F_{\bar\nu_e}}{F_{\bar\nu_e}-F_{\bar\nu_x}}\,.
\end{equation}
Eq.~(\ref{eq:energy}) then implies that the ``flavor pendulum'' is in a stable initial position
for the normal hierarchy, $\theta\ll1$, whereas for the inverted hierarchy
$\tilde\theta=\pi/2-\theta\ll1$ the pendulum is initially in a maximum energy state.
We will focus on this case in the following.

As long as $\mu>2\omega/(1-\sqrt{1+\varepsilon})^2$ the kinetic (self-interaction)
term in Eq.~(\ref{eq:energy}) dominates and the oscillations are
synchronized with a common frequency $\omega_{\rm synch}=(2+\varepsilon)\omega/\varepsilon$
around ${\bf B}$. For $\omega<\mu<2\omega/(1-\sqrt{1+\varepsilon})^2$, bipolar oscillations
take place with a frequency $\kappa=(2\omega\mu/(1+\varepsilon))^{1/2}$ in which ${\bf Q}$
swings between its initial position and a position in which it is parallel to $-{\bf B}$.
This corresponds to a collective transition of $\nu_e\bar\nu_e$ to $\nu_x\bar\nu_x$ pairs,
with a rate greatly speeded up compared to ordinary pair annihilation~\cite{Sawyer:2005jk}.
Furthermore, in the absence of matter ${\bf D}\cdot{\bf B}=({\bf P}-{\bf\bar P})\cdot{\bf B}$
is strictly conserved~\cite{Hannestad:2006nj}, whereas in dense matter $({\bf P}-{\bf\bar P})\cdot{\bf L}$ is
approximately conserved. For small effective mixing angle, one thus has $P_z=\bar P_z+\epsilon$.
Finally, when $\mu<\omega$ at large radii, vacuum oscillations ensue or, in the presence
of matter with $\lambda\sim\omega$ the usual MSW effects can occur.

In the limit of small vacuum mixing angle, $\tilde\theta=\pi/2-\theta\ll1$, and for constant
$\mu$ and $\lambda$, the time scale for bipolar conversion is
\begin{equation}\label{eq:tau_bipolar}
 \tau_{\rm bipolar}\simeq-\kappa^{-1}\,
 \ln\left(\frac{\tilde\theta\kappa}{(\kappa^2+\lambda^2)^{1/2}}\right)\,.
\end{equation}
The main effect of the matter term is thus to decrease the effective
mixing angle to $\tilde\theta\kappa/(\kappa^2+\lambda^2)^{1/2}$ and thus delay
the onset of bipolar oscillations. For smaller mixing angle it thus takes
longer to "tip over", the bipolar transition is less adiabatic and the
nutation amplitude is larger. Since matter decreases the effective mixing
angle, a high matter density also leads to a later onset of the bipolar transition
and to a larger nutation amplitude~\cite{EstebanPretel:2007ec}.

For varying neutrino density and in the limit of small vacuum mixing
angle, angular momentum and adiabatic energy conservation in the
bipolar regime gives for the electron neutrino survival probability
$P(\nu_e\to\nu_e)=\frac{1}{2}(1+P_z)\propto\mu(r)^{1/2}$.

Since the solar neutrino mass hierarchy is known to be normal, the bipolar conversion in
the inverted atmospheric hierarchy is essentially pair conversion
$\nu_e\bar\nu_e\to\nu_x\bar\nu_x$ from the second highest state $m_1$ to the lowest
state $m_3$, where $\nu_x$ is a combination of $\nu_\mu$ and $\nu_\tau$.
We stress that for the inverted hierarchy, bipolar oscillations occur
for arbitrarily small $\theta_{13}$. This can be used as experimental test by observing
a supernova with mega-ton detectors that are mostly sensitive to electron-antineutrinos
via the reaction $\bar\nu_e+p\to n+e^+$~\cite{Dasgupta:2008my}: If either the atmospheric
hierarchy is normal or if it is inverted with $\sin^2\theta_{13}\ga10^{-3}$, the
bipolar transition is followed by an adiabatic MSW transition, and the electron
anti-neutrino fluxes seen in two detectors behind and in front of the Earth
are different. In contrast, if the atmospheric hierarchy is inverted and $\sin^2\theta_{13}\la10^{-5}$,
the bipolar transition is followed by a non-adiabatic MSW transition, and the electron
anti-neutrino fluxes seen in two such detectors are equal.

\section{Kinematic and Self-Induced Decoherence between Angular Modes}
The momentum modes ${\bf p}$ essentially consist of energy and angular modes. For
spherical symmetry one has the radius $r$ as integration variable, such that
\begin{eqnarray}\label{eq:modes}
 {\bf p}&\to&(E\simeq|{\bf p}|, u\equiv\sin^2\theta_R)\,,\\
 \varrho_{\bf p}(r)&\to&\varrho_{E,u,r}\,,\nonumber
\end{eqnarray}
where $\theta_R$ is the emission angle relative to the radial direction at the neutrino sphere,
$r=R$. The energy bins
will lead to spectral splits~\cite{Duan:2006an,Raffelt:2007cb,Duan:2007bt,Fogli:2007bk},
whereas angular bins can give rise to kinematic and self-induced
decoherence~\cite{Raffelt:2007yz,EstebanPretel:2007ec}. To
see this, we define the flux matrices
\begin{equation}\label{eq:flux_matrix}
 J_{E,u,r}\equiv\frac{E^2\varrho_{E,u,r}}{2(2\pi)^2}\,,
\end{equation}
With the radial velocity $v_{u,r}=v_{{\bf p},r}=\cos\theta_r=\sqrt{1-u(R/r)^2}$ one has
the integral flux and number density matrices
\begin{eqnarray}\label{eq:flux_matrices}
  J_r&\equiv&\frac{r^2}{R^2}\int\frac{\D^3{\bf p}}{(2\pi)^3}\,\varrho_{\bf p}=
  \int_0^1 du\int_0^\infty dE\,J_{E,u,r}\,,\nonumber\\
  N_r&\equiv&\int_0^1 du\int_0^\infty dE\frac{J_{E,u,r}}{v_{u,r}}\,.
\end{eqnarray}
For an average matter velocity $v_e$ one then has the general equations of motion:
\begin{eqnarray}\label{eq:eom2}
 i\partial_rJ_{E,u,r}&=&\biggl[\frac{\Omega_E+\lambda(r)L(1-v_{u,r}v_e)}{v_{u,r}}\\
 &&\hskip-2cm+\sqrt{2}\,G_{\rm F}\frac{R^2}{r^2}
 \left(\frac{N_r-\bar N_r}{v_{u,r}}-(J_r-\bar J_r)\right),J_{E,u,r}\biggr]
 \,,\nonumber
\end{eqnarray}
and analogously for antineutrinos with the opposite sign for $\Omega_E$.
Decoherence can occur due to the the $u-$dependence of the velocity $v_{u,r}$,
either from the flux term in the self-interactions (self-induced decoherence) or
kinematically in the matter terms. The matter term can be transformed away up to the
$u-$dependence of $v_{u,r}$~\cite{Duan:2005cp,Hannestad:2006nj}. The matter-induced
multi-angle effect becomes important when
\begin{equation}\label{eq:multiangle_matter}
 n_e=n_{e^-}-n_{e^+}\ga n_{\bar\nu_e}\,.
\end{equation}
If the matter density is very much larger than the neutrino density, the effective
oscillation frequencies of different polarization vectors vary so greatly that they
stay pinned to the ${\bf L}$ direction and no collective oscillations occur.

Numerical simulations have been performed to determine under which conditions decoherence
occurs. For $\omega=0.3~{\rm km}^{-1}$, $\sin2\tilde\theta=10^{-3}$ and $\sin2\theta=10^{-3}$, the
demarcation lines between coherence and decoherence for inverted and normal hierarchies,
respectively, in the $\mu$-$\epsilon$-plane, can be approximated by~\cite{EstebanPretel:2007ec}
\begin{eqnarray}\label{eq:epscontour}
 \epsilon_{\rm IH}&\approx&0.225+0.027\,\log_{10}
 \left(\frac{\mu}{10^6~{\rm km}^{-1}}\right)\,,
 \nonumber\\*
 \epsilon_{\rm NH}&\approx&0.172+0.087\,\log_{10}
 \left(\frac{\mu}{10^6~{\rm km}^{-1}}\right)\,.
\end{eqnarray}

\section{Energy Modes and Spectral Splits}
Integrating Eq.~(\ref{eq:eom2}) over $E$ and $u$ and writing for the two flavor case
$J_r=\left[F_\nu-F_{\bar\nu}+{\bf D}\cdot\bm\sigma\right]/2$, in the limit of small
mixing angles one gets~\cite{Raffelt:2007cb}
\begin{equation}\label{eq:integraleom}
 \partial_r{\bf D}={\bf e}_z\times{\bf M}\,,
\end{equation}
where ${\bf M}$ is an integral of $J_{E,u,r}$ which is not interesting for our purposes.
Thus, $(J_r-\bar J_r)_{22}-(J_r-\bar J_r)_{11}=F_{\nu_e}-F_{\bar\nu_e}-\left[F_{\nu_x}-F_{\bar\nu_x}\right]=\,$const,
which corresponds to the approximate conservation of $({\bf P}-{\bf\bar P})\cdot{\bf L}$ in
the single mode approximation.
Since the total lepton number is conserved, electron and $x-$lepton numbers are conserved
separately.

The spectral split is governed by lepton number conservation for both flavors
separately: Anti-neutrinos swap completely in a bipolar transition. To compensate, neutrinos
can only swap above a certain energy because under typical supernova conditions
\begin{equation}\label{eq:swap}
|F_{\nu_x}-F_{\nu_e}|>|F_{\bar\nu_x}-F_{\bar\nu_e}|\,.
\end{equation}
The spectral splits are created by an adiabatic transition between the regime dominated
by neutrino-self interactions and the low neutrino density regime~\cite{Raffelt:2007cb}.
If this transition is not completely adiabatic, as may be the case in a real supernova,
the spectral split tends to be washed out~\cite{Raffelt:2007xt}. Spectra splits may be
observable in future observations of a galactic supernova explosion~\cite{Fogli:2008fj}.
Such splits may also occur in the antineutrino sector~\cite{Fogli:2007bk,Fogli:2008pt},
albeit at lower energies and it is not clear if such features are not washed out when
taking into account angular modes~\cite{Fogli:2008pt}.

\section{Three-flavor Effects}\label{sec4}

\begin{figure}[h!]
\includegraphics*[width=0.5\textwidth]{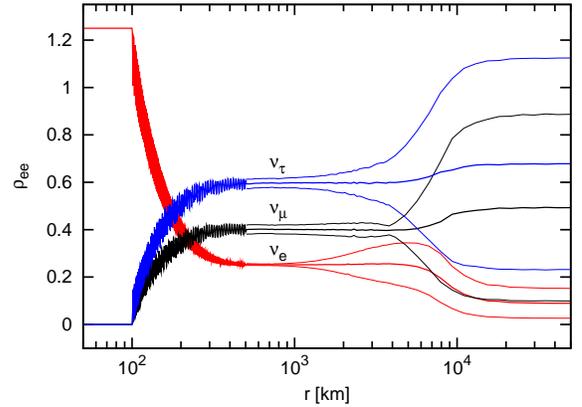}
\caption{Radial evolution of the fluxes of $\nu_e$ (red), $\nu_\mu$ (black) and
$\nu_\tau$ (blue) fluxes, normalized to the initial $\bar\nu_e$-flux,
for a fixed neutrino energy $E=20\,$MeV and an inverted atmospheric hierarchy.
For the neutrino parameters we use $\Delta m^2_{12}=\Delta m^2_{\rm
sol}=7.65\times10^{-5}~{\rm eV}^2$, $\Delta m^2_{13}=\Delta m^2_{\rm
atm}=2.4\times10^{-3}~{\rm eV}^2$, $\sin^2\theta_{12}=0.304$,
$\sin^2\theta_{13}=0.01$, $\sin^2\theta_{23}=0.4$, and a vanishing Dirac phase
$\delta=0$, all consistent with measurements~\cite{Schwetz:2008er}.
The matter density profile $\lambda(r)=4\times10^6\,(R/r)^3{\rm km}^{-1}$
was assumed, where the neutrinosphere is at $R=10\,$km. The self-interaction term is
taken as $\mu(r)=7\times10^5\,(R/r)^4/(2-(R/r)^2){\rm km}^{-1}$. After bipolar conversion,
thick lines represent the average fluxes, whereas thin lines signify the envelopes
of the fast flux oscillations.}
\label{fig1}
\end{figure}

\begin{figure}[h!]
\includegraphics*[width=0.5\textwidth]{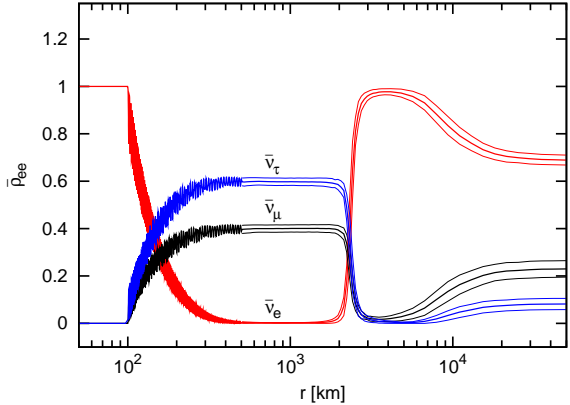}
\caption{Same as Fig.~\ref{fig1}, but for the antineutrino fluxes.}
\label{fig2}
\end{figure}

In normal matter, $\mu$ and $\tau$ leptons appear only as virtual
states in radiative corrections to neutral-current $\nu_\mu$ and
$\nu_\tau$ scattering, causing a shift $\Delta V_{\mu\tau}=
\sqrt{2}\,G_{\rm F}Y_\tau^{\rm eff}n_B$ between $\nu_\mu$
and~$\nu_\tau$, where $n_B$ is the baryon density. It has the same
effect on neutrino dispersion as real $\tau$ leptons with an abundance~\cite{Botella:1986wy}
\begin{eqnarray}\label{eq:Ytau}
 Y_\tau^{\rm eff}&=&\frac{3\sqrt{2}\,G_{\rm F}m_\tau^2}{(2\pi)^2}
 \left[\ln\left(\frac{m_W^2}{m_\tau^2}\right)-1+\frac{Y_n}{3}\right]\nonumber\\
 &\simeq&2.7\times10^{-5}\,,
\end{eqnarray}
where $n_e$ was assumed to equal the proton density and $Y_n$ is the neutron
fraction of $n_B$. This should be compared to the ordinary
MSW potential $\Delta V=\sqrt{2}\,G_{\rm F}Y_e n_B$ where
$Y_e=n_e/n_B$ is the electron fraction.

The potential Eq.~(\ref{eq:Ytau}) would lead to a MSW resonance at density
$\rho\simeq3\times10^7\,{\rm g}\,{\rm cm}^{-3}$, corresponding
to $\lambda\simeq10^4\,{\rm km}^{-1}$, provided that the radius at which
this resonance occurs is beyond the radii at which collective oscillations occur.
The $\nu_e$ and $\bar\nu_e$ survival probabilities would then be sensitive to
the $\theta_{23}$ angle which governs $\nu_\mu-\nu_\tau$ mixing and such
a situation could a priori arise in the accretion phase of an iron-core
supernova~\cite{EstebanPretel:2007yq}.

However, the $\nu_\mu-\nu_\tau$ refractive effect is
unlikely to play any practical role: At the required high matter densities
the multi-angle matter effect is likely to trigger multi-angle decoherence such that
the fluxes of the different flavors tend to be maximally equilibrated~\cite{EstebanPretel:2008ni}.

In Figs.~\ref{fig1} and~\ref{fig2} we show the radial dependence of neutrino and
antineutrino fluxes for a typical example of three-flavor oscillations in
single mode approximation.
The asymmetry parameter $\varepsilon=0.25$ from Eq.~(\ref{eq:epsdefine}) is
sufficiently large to prevent self-induced decoherence according to Eq.~(\ref{eq:epscontour}).
Since $\mu(r)\ga\lambda(r)(1-v_{u,r})\simeq\lambda(r)(R/r)^2/2$ outside the
synchronized oscillation regime, matter induced decoherence should also be
negligible according to Eq.~(\ref{eq:multiangle_matter}). Only one radial mode
was thus taken into account in this simulation. After
the synchronized oscillation phase which lasts until $r\simeq100\,$km, a
bipolar transition occurs which lasts until $r\simeq300\,$km. During this
bipolar phase, the survival probability of electron antineutrinos falls
off roughly as $\mu(r)^{1/2}$, as discussed in Sect.~\ref{sec2}, whereas the electron
neutrino survival probability approaches the value $\varepsilon=0.25$, as
dictated by approximate flavor conservation. If energy modes would be included
in such a simulation, this would result in a spectral split such that
the $\nu_e$ flux would not swap with the $\nu_\mu+\nu_\tau$ fluxes below
a certain critical energy determined by approximate flavor conservation. Furthermore, a
$\nu_\mu-\nu_\tau$ MSW transition would occur at $r\simeq100\,$km, were it not
for the collective effects that prevent such a transition. Finally, an MSW
resonance occurs at $r\simeq2500\,$km for antineutrinos due to the assumed
inverted hierarchy, after which vacuum oscillations remain.

\section{Open Questions and Conclusions}
We first summarize our main conclusions.
Neutrino self-interactions can play a major role in the oscillations of neutrinos
in core collapse supernovae.
The one-mode approximation is meanwhile well understood and can be
thought of as a spinning top. It often leads to a surprisingly accurate
description of the full problem which in the simplest case of spherical symmetry
requires the introduction of modes both for energy and for the
direction of a given neutrino trajectory with respect to the radial direction.
Asymmetry between neutrinos and anti-neutrinos leads to spectral splits
in the context of energy modes and, if sufficiently large, prevents
self-induced decoherence of angular modes.
Matter effects can also lead to decoherence if the charged lepton density
is larger than the neutrino density. This is also likely to mask any significant
effects of the second-order difference between the $\nu_\mu$ and $\nu_\tau$
refractive index on resulting neutrino fluxes. If matter densities are very much
larger than neutrino densities, the multi-mode flavor polarization vectors
remain pinned to the direction corresponding to flavor eigenstates and no
flavor conversion occurs.

There are still unresolved issues in collective neutrino oscillations, including
their detailed numerical description in the absence of spherical symmetry, which can
be relevant, for example, in the presence of hydrodynamic turbulence in the
background of ordinary matter. It is, for example, currently not
completely clear, even conceptually, how to describe damping due to the different
matter profiles ``seen'' along different neutrino trajectories in this context.
Another interesting question could be if their could be any significant dependence
of collective oscillations on the Dirac phase $\delta$~\cite{Gava:2008rp}.


\section*{Acknowledgments}
This work was supported by the Deutsche Forschungsgemeinschaft (SFB 676 ``Particles, Strings
and the Early Universe: The Structure of Matter and Space-Time) and by the European Union
(contracts No. RII3-CT-2004-506222).



\bibliographystyle{aipprocl} 

\begin{thebibliography}{9}

\bibitem{Sawyer:2005jk}
  R.~F.~Sawyer,
  Phys.\ Rev.\  D {\bf 72}, 045003 (2005)
  [hep-ph/0503013].

\bibitem{Duan:2005cp}
  H.~Duan, G.~M.~Fuller and Y.~Z.~Qian,
  Phys.\ Rev.\  D {\bf 74}, 123004 (2006)
  [arXiv:astro-ph/0511275].

\bibitem{Duan:2006an}
H.~Duan, G.M.~Fuller, J.~Carlson and Y.Z.~Qian,
Phys.\ Rev.\ D {\bf 74}, 105014 (2006)
[arXiv:astro-ph/0606616].

\bibitem{Hannestad:2006nj}
  S.~Hannestad, G.~G.~Raffelt, G.~Sigl and Y.~Y.~Y.~Wong,
  Phys.\ Rev.\ D {\bf 74}, 105010 (2006)
  [astro-ph/0608695].

\bibitem{Raffelt:2007yz}
  G.~G.~Raffelt and G.~Sigl,
  Phys.\ Rev.\  D {\bf 75}, 083002 (2007)
  [arXiv:hep-ph/0701182].

\bibitem{EstebanPretel:2007ec}
  A.~Esteban-Pretel, S.~Pastor, R.~Tomas, G.~G.~Raffelt and G.~Sigl,
  Phys.\ Rev.\  D {\bf 76}, 125018 (2007)
  [arXiv:0706.2498 [astro-ph]].

\bibitem{Raffelt:2007cb}
G.G.~Raffelt and A.Yu.~Smirnov,
Phys.\ Rev.\  D {\bf 76}, 081301 (2007)
[arXiv:0705.1830].

\bibitem{Raffelt:2007xt}
  G.~G.~Raffelt and A.~Y.~Smirnov,
  Phys.\ Rev.\  D {\bf 76}, 125008 (2007)
  [arXiv:0709.4641 [hep-ph]].

\bibitem{Duan:2007bt}
 H.~Duan, G.M.~Fuller, J.~Carlson and Y.Z.~Qian,
Phys.\ Rev.\ Lett.\  {\bf 99}, 241802 (2007)
[arXiv:0707.0290].

\bibitem{Fogli:2007bk}
 G.L.~Fogli, E.~Lisi, A.~Marrone and A.~Mirizzi,
J.\ Cosmol.\ Astropart.\ Phys.\ {\bf 12}, 010 (2007)
[arXiv:0707.1998].

\bibitem{EstebanPretel:2007yq}
  A.~Esteban-Pretel, S.~Pastor, R.~Tomas, G.~G.~Raffelt and G.~Sigl,
  Phys.\ Rev.\  D {\bf 77}, 065024 (2008)
  [arXiv:0712.1137 [astro-ph]].

\bibitem{EstebanPretel:2008ni}
  A.~Esteban-Pretel, A.~Mirizzi, S.~Pastor, R.~Tomas, G.~G.~Raffelt, P.~D.~Serpico and G.~Sigl,
  Phys.\ Rev.\  D {\bf 78}, 085012 (2008)
  [arXiv:0807.0659 [astro-ph]].

\bibitem{Fogli:2008pt}
  G.~L.~Fogli, E.~Lisi, A.~Marrone, A.~Mirizzi and I.~Tamborra,
  Phys.\ Rev.\  D {\bf 78}, 097301 (2008)
  [arXiv:0808.0807 [hep-ph]].

\bibitem{Fogli:2008fj}
  G.~Fogli, E.~Lisi, A.~Marrone and I.~Tamborra,
  arXiv:0812.3031 [hep-ph].

\bibitem{Dolgov:1980cq}
  A.~D.~Dolgov,
  ``Neutrinos in the early universe,''
  Yad.\ Fiz.\  {\bf 33}, 1309 (1981)
  [Sov.\ J.\ Nucl.\ Phys.\  {\bf 33}, 700 (1981)].

\bibitem{Sigl:1992fn}
  G.~Sigl and G.~Raffelt,
  ``General kinetic description of relativistic mixed neutrinos,''
  Nucl.\ Phys.\ B {\bf 406}, 423 (1993).

\bibitem{mckellar&thomson}
  B.~H.~J.~McKellar and M.~J.~Thomson,
  ``Oscillating doublet neutrinos in the early universe,''
  Phys.\ Rev.\ D {\bf 49}, 2710 (1994).

\bibitem{Wolfenstein:1977ue}
L.~Wolfenstein, ``Neutrino oscillations in matter,''
{{\em Phys. Rev.} {\bf D17}
  (1978)  2369--2374}.

\bibitem{Mikheyev:1985aa}
S.~P. Mikheyev and A.~Y. Smirnov {\em Yad. Fiz.} (1985) no.~42, 1441. [Sov.\
  J.\ Nucl.\ Phys.\ {\bf 42}, 913 (1985)].

\bibitem{Schwetz:2008er}
  see, e.g., T.~Schwetz, M.~Tortola and J.~W.~F.~Valle,
  New J.\ Phys.\  {\bf 10}, 113011 (2008)
  [arXiv:0808.2016 [hep-ph]].

\bibitem{Dasgupta:2008my}
  B.~Dasgupta, A.~Dighe and A.~Mirizzi,
  Phys.\ Rev.\ Lett.\  {\bf 101}, 171801 (2008)
  [arXiv:0802.1481 [hep-ph]].

\bibitem{Botella:1986wy}
F.J.~Botella, C.S.~Lim and W.J.~Marciano,
Phys.\ Rev.\  D {\bf 35}, 896 (1987).

\bibitem{Gava:2008rp}
  J.~Gava and C.~Volpe,
  Phys.\ Rev.\  D {\bf 78}, 083007 (2008)
  [arXiv:0807.3418 [astro-ph]].

\end{thebibliography}

\end{document}